\documentclass[aps,prc,reprint,superscriptaddress]{revtex4-1}

\usepackage{graphicx}
\graphicspath{{./figs/}}

\usepackage[english]{babel}
\usepackage[reqno,sumlimits,intlimits]{amsmath}
\usepackage{tabularx}
\usepackage{array} 
\newcolumntype{M}[1]{>{\centering\arraybackslash}m{#1}} 
\usepackage[caption=false]{subfig}
\usepackage[breaklinks]{hyperref}

\begin{document}
\title{Hadronic vs. partonic $\boldsymbol{J/\psi}$ production in the Statistical Hadronization Model}

\author{Philipp B. Bahavar}
\email{bahavar@th.physik.uni-frankfurt.de}
\author{Jan Uphoff}
\affiliation{Institut f\"ur Theoretische Physik, Johann Wolfgang 
Goethe-Universit\"at Frankfurt, Max-von-Laue-Str.~1, \\
D-60438 Frankfurt am Main, Germany}
\author{Carsten Greiner}
\affiliation{Institut f\"ur Theoretische Physik, Johann Wolfgang 
Goethe-Universit\"at Frankfurt, Max-von-Laue-Str.~1, \\
D-60438 Frankfurt am Main, Germany}

\date{May 8, 2014}

\begin{abstract}
Assuming the Statistical Hadronization Model and charm flavour conservation, one can quantify the deviation of charm and anti-charm quarks from chemical equilibrium both in the hadronic phase and in the quark-gluon plasma. By linking the hadronic fireball volume and the corresponding plasma source volume via entropy conservation as proposed by Grandchamp \emph{et al.}\cite{Grandchamp:2003uw}, a direct comparison between the $J/\psi$ yields in both pictures can be used to obtain limits for the charm quark mass in the medium.
\end{abstract}

\maketitle
\section{Introduction}
Experimental results from ultra-relativistic heavy ion collisions at the BNL Relativistic Heavy Ion Collider (RHIC) \cite{Star05,Phenix05,Brahms05,Phobos05} and at the CERN Large Hadron Collider (LHC) \cite{LHC12} suggest the formation of a new state of strongly interacting matter, the quark-gluon plasma (QGP). The Statistical Hadronization Model (SHM) has been very successful in describing the yields of hadrons with light valence quarks ($u$, $d$, $s$) \cite{PBM01,And05} obtained at various collision energies in these experiments, but did originally not include charmed hadrons and charmonia. Since $J/\psi$ suppression was proposed as a distinctive signature of such a hot and dense partonic medium \cite{MS86}, models describing charmonium production have been of special interest in high energy nuclear physics. Including open and hidden charm hadrons by explicitly fixing the number of produced $c\bar{c}$-pairs and adjusting the hadron yields according to flavour conservation \cite{rev09} extended the SHM towards this goal and proved to be able to reproduce also the yields of hadrons containing charm quarks \cite{And03}. As calculations on the lattice show $J/\psi$ dissociate only at temperatures between $1.5$ and $1.9$ times the critical temperature $T_c$ \cite{latticeQCD03,Karsch03,latticeQCD07}, further extending the model to include the behaviour in the plasma phase is desirable.\\

In this letter we briefly recall the formalism of the SHM with charm quarks (Section \ref{sec:Statistical Hadronization Model}), and then argue in Section \ref{sec:Charm fugacity in the QGP} that this approach can be adapted to describe not only the $J/\psi$ in the vicinity of the hadronic phase, but also charm and anti-charm quarks in the QGP phase before hadronization to account for the aforementioned lattice results following the proposal from \cite{Grandchamp:2003uw}. Assuming the $J/\psi$ mesons equilibrate with the medium under the constraint of detailed charm balance (e.g. $J/\psi + g \leftrightarrow c + \bar{c} $), compatible yields are expected in both descriptions at the freeze-out temperature, as the binding energy of the $J/\psi$ should be sufficient to survive any phenomena occurring during hadronization of the medium. We will show in Section \ref{sec:Comparison of multiplicities} that this imposes stringent constraints on the possible effective dressed mass of the charm quark. 
\section{Statistical Hadronization Model}
\label{sec:Statistical Hadronization Model}
The Statistical Hadronization Model (SHM) provides a framework in which the particle distribution across the different hadron states and resonances is described by thermodynamic quantities associated with the fireball that exists in the aftermath of a heavy ion collision. In the SHM, the source medium is assumed to be in complete thermal and chemical equilibrium, and the hadronic states form a grand-canonical ensemble described by the partition function $Z^{\mathrm{GC}}$, leading to \citep{And05} 
\begin{align}
\label{eq:partitionFunction}
\ln Z^{\mathrm{GC}}_i=\pm \frac{Vg_i}{2\pi^2}\int_{0}^{\infty}\!\mathrm{d}p\, p^2\mathrm{ln}\left(1\pm e^{-\frac{E_i-\mu_i}{T}}\right),
\end{align}
where $g_i$ is the degeneracy factor due to spin, $E_i=\sqrt{p^2+m_i^2}$ is the total energy, and $\mu_i=B_i\mu_B+S_i\mu_S+C_i\mu_C+I_{3_i}\mu_{I_{3_i}}$ is the chemical potential associated with the quantum charges of particle species $i$. The upper sign applies to fermions, the lower to bosons. For numerical purposes, a series representation of this expression can be used as follows \cite{PBM03}:
\begin{align}
\label{eq:partitionSeries}
\ln Z^{\mathrm{GC}}_i=\frac{VTg_i}{2\pi^2}\,\sum_{n=1}^\infty\frac{(\mp 1)^{n+1}}{n^2} \lambda_i^n m_i^2 K_2\left(\frac{nm_i}{T}\right),
\end{align}
with the modified Bessel function of the second kind $K_2$ and  the same sign convention as above. The fugacity is defined as $\lambda_i=\exp\left(\mu_i/T\right)$. Because of the suppression of higher order terms by $\exp\left(-m/T\right)$, the Boltzmann description given by the first term is sufficient for all but the lightest hadrons, i.e. pions. From here, particle multiplicities can be obtained via \cite{And05}
\begin{align}
\label{eq:multiplicity}
\begin{split}
&N_i^{\mathrm{th}}(T,\mu) = T\,\frac{\partial\ln Z^{\mathrm{GC}}_i}{\partial \mu}\\
&=\frac{VTg_i}{2\pi^2}\,\sum_{n=1}^\infty (\mp 1)^{n+1}\frac{m_i^2}{n}\cdot\lambda_i^n(T,\mu)\cdot K_2\left(\frac{nm_i}{T}\right).
\end{split}
\end{align}
The chemical freeze-out temperature $T$ and the baryo-chemical potential $\mu_B$ in particular are enough to determine particle ratios within this description, and can in turn be calculated from fitting the experimentally observed ratios, assuming that both parameters are universal for all hadron species. The hadronization volume $V_{\mathrm{SHM}}\equiv \mathrm{d}V/\mathrm{d}y\lvert_{\Delta y=1}$ is obtained using the number of charged hadrons measured at midrapidity $N_{\mathrm{ch}}\equiv \mathrm{d}N_{\mathrm{ch}}/\mathrm{d}y\lvert_{\Delta y=1}$ as input and calculating the corresponding expected density $n_\mathrm{ch}$ within the SHM. Because all multiplicities $N$ are calculated using this midrapidity volume slice, $N\equiv\mathrm{d}N/\mathrm{d}y\lvert_{\Delta y=1}$ is used universally in the following.
\begin{figure}[htb]
\centering
\includegraphics[width=\linewidth]{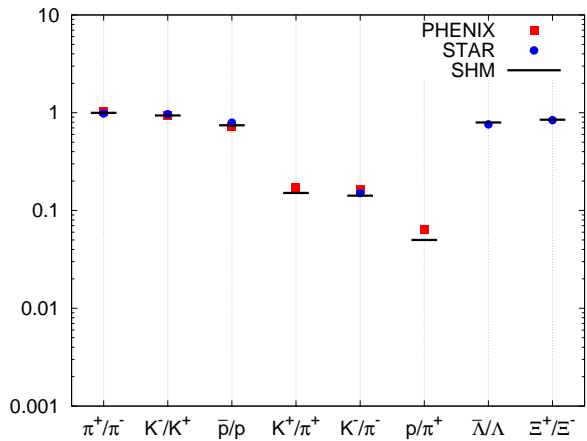}
\caption{Particle ratios as reported by \cite{PHENIX,STAR} reproduced by the SHM with $T=162\,\mathrm{MeV}$ and $\mu_B=24\,\mathrm{MeV}$, as suggested in \cite{Andronic:2012dm}.}
\label{fig:particle_ratios}
\end{figure}
Additionally, corrections for the finite volume of the hadrons, which were up to this point treated as point particles, have to be made by modifying the chemical potential \cite{Rischke91,YG97}. Considering a hadron gas containing $k$ different particle species $i$ with excluded volume $v_i$, iteration of
\begin{align}
\label{eq:p-mu-iteration}
\begin{aligned}
\hat{p}\left(T,\mu\right)=p^{\mathrm{id}}\left(T,\hat{\mu}_1,\ldots,\hat{\mu}_k\right)\\
\hat{\mu}_i=\mu_i-v_i\cdot \hat{p}\left(T,\mu_1,\ldots,\mu_k\right)
\end{aligned}
\end{align}
with the pressure 
\begin{align}
\label{eq:pressure}
\begin{split}
&p^{\mathrm{id}}_i(T,\mu)=T\frac{\partial\ln Z^{\mathrm{GC}}_i}{\partial V}\\
&=\frac{g_iT^2}{2\pi^2}\,\sum_{n=1}^\infty\frac{(\mp 1)^{n+1}}{n^2}\lambda_i^n(T,\mu)\cdot m^2\cdot K_2\left(\frac{nm_i}{T}\right)
\end{split}
\end{align}
yields a new chemical potential $\hat{\mu}$, which is then used in the various expressions derived before. Thermal densities are then further reduced by a factor of 
\begin{align}
\frac{1}{1+\sum_i v_i n_i^{\mathrm{th}}(T,\hat{\mu})}.
\end{align}
The excluded volume of the hadrons was taken to be $v_i\equiv v = 16/3\pi R^3$ with $R=0.3\,\mathrm{fm}$ for all species reported by the Particle Data Group \cite{PDG}. Figure \ref{fig:particle_ratios} shows the well-known very good agreement of this implementation of the Statistical Hadronization Model in comparison to particle ratios measured by the PHENIX and STAR collaborations.\\
 
While the Statistical Hadronization Model originally assumes hadron production into perfect equilibrium, it can be extended to include hadron states which do not fulfill this condition. This procedure is necessary to describe hadrons containing charm or anti-charm quarks in ultra-relativistic heavy-ion collisions \cite{PBM03,And03,And07}, which are produced away from the standard chemical equilibrium of open charm \cite{Uphoff10}. Accordingly, these states are treated via fixing the total available number of $c\bar{c}$-pairs in the system. Because of the high mass of the charm quark $m_c$ compared to the thermal energy scale at freeze-out, we assume this number to be approximately equal to the number of pairs produced in the initial hard collision $N_{c\bar{c}}^{\mathrm{dir}}$, which can be calculated using the charm production cross-section for heavy ion collisions \cite{rev09}. As a side remark see, however, \cite{Uphoff10} for an estimation of secondary charm production in the medium: These
secondary production processes are small at RHIC and might be moderate at LHC energies depending on the initial conditions
for the gluons. The thermal yields for charmed hadrons $N_{\mathrm{oc}}^{\mathrm{th}}$ and charmonia $N_{c\bar{c}}^{\mathrm{th}}$ are then amplified by the charm fugacity parameter $g_c$ to satisfy exact charm conservation as expressed in the charm balance equation \cite{And03}:
\begin{align}
\label{eq:charm balance equation}
N_{c\bar{c}}^{\mathrm{dir}}=\frac{1}{2} g_c N_{\mathrm{oc}}^{\mathrm{th}} \frac{I_1\left(g_c N_{\mathrm{oc}}^{\mathrm{th}}\right)}{I_0\left(g_c N_{\mathrm{oc}}^{\mathrm{th}}\right)} + g_c^2 N_{c\bar{c}}^{\mathrm{th}}.
\end{align}
This leads to a charmionium enhancement by a factor of $g_c^2$ compared to the expectation for purely thermal production. The ratio of the modified Bessel functions $I_1/I_0$  accounts for canonical suppression effects \cite{Foc06} caused by the possibly small number of charm quarks in the system and approaches unity in the grand-canonical limit.
\section{Charm fugacity in the QGP}
\label{sec:Charm fugacity in the QGP}
In principle, the SHM can not take into account any dynamical processes which happen in a heavy-ion-collision prior to hadronization. Nevertheless, using knowledge about the composition and properties of the QGP, it is possible to translate some of the phenomena discussed into the partonic sector of the phase diagram as follows:\\
As the phase transition  or crossover transition happens on a short time scale and if, as assumed, hadrons are produced into equilibrium, it stands to reason  that an equilibrated partonic medium existed just before the hadronization process. It can further be argued that the deviation of charmed hadron and charmonium multiplicities from the values for chemical equilibrium quantified by the charm fugacity factor $g_c$ is a remnant of the corresponding deviation of charm quark numbers from the equilibrium expectation value in the QGP. Analogous to the approach in the hadronic sector described by the SHM, a charm fugacity $\lambda_c$ can also be introduced and calculated on the quark level in the QGP prior the hadronization.\\

Assuming the entropy $S$ to be constant across the hadronization process, a QGP volume $V_{\mathrm{QGP}}$ can be obtained using the expression for the entropy density $s_{\mathrm{QGP}}$ of the QGP \cite{Rischke03}
\begin{align}
\label{eq:V_QGP}
V_{\mathrm{QGP}}=\frac{S_{\mathrm{SH M}}}{s_{\mathrm{QGP}}}= S_{\mathrm{SHM}} \left[4\left(16+\frac{21}{2}N_f\right)\frac{\pi^2}{90}\,T^3\right]^{-1}.
\end{align}
$N_f$ corresponds to the number of massless flavours in the QGP and is set to $N_f=3$, as the charm quark is explicitly treated as a massive particle. The entropy of one particle species $i$ in the hadron gas
\begin{align}
S_{i}^{\mathrm{SHM}}=\frac{g_i V_{\mathrm{SHM}}}{2\pi^2}m_i^{2}\sum_{n=1}^{\infty}\left(\mp1\right)^{n+1}\lambda_i^n\frac{\mathcal{G}_n(T,\mu)}{n^{2}}
\end{align}
with
\begin{align}
\begin{split}
\mathcal{G}_n(T,\mu)=&\left(2T-\mu n\right) K_2\left(\frac{nm_i}{T}\right)\\&+\frac{nm_i}{2}\left[K_1\left(\frac{nm_i}{T}\right)+K_3\left(\frac{nm_i}{T}\right)\right]
\end{split}
\end{align}
follows directly from Equation (\ref{eq:partitionSeries}) via
\begin{align}
S_{i}^{\mathrm{SHM}}=\frac{\partial}{\partial T}\left(T\ln Z^{\mathrm{GC}}_i\right),
\end{align}
taking into account both the excluded volume correction and feed-down from higher hadronic states.\\

\begin{figure}[htb]
\centering
\includegraphics[width=\linewidth]{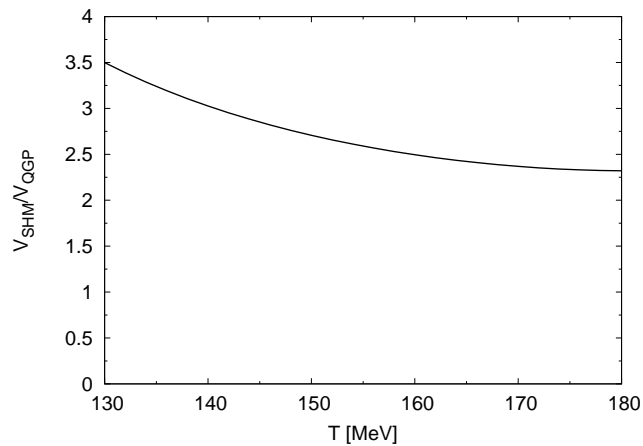}
\caption{The ratio between the fireball volume $V_{\mathrm{SHM}}$ and the plasma volume $V_{\mathrm{QGP}}$ is shown as a function of the assumed freeze-out temperature.}
\label{fig:volume_scaling}
\end{figure} 
The relation between $V_{\mathrm{SHM}}$ and the resulting $V_{\mathrm{QGP}}$ is shown in Figure \ref{fig:volume_scaling}. The QGP volume is calculated to be smaller than the hadronic fireball volume by a factor of $2.5$ at $T=160\,\mathrm{MeV}$. Due to the finite hadron volumes, the ratio does not approach unity for high temperatures.\\

Now the charm fugacity is equal to the ratio of $N_{c\bar{c}}^{\mathrm{dir}}$ to the number of charm quarks expected in equilibrium $N_c^{\mathrm{th}}$:
\begin{align}
\label{eq:lambda_c}
N_{c\bar{c}}^{\mathrm{dir}}=\lambda_c \frac{2 V_{\mathrm{QGP}}}{2\pi^2}\left(m_cT\right)^{\frac{3}{2}}\exp\left(-m_c/T\right),
\end{align}
neglecting effects from the baryochemical potential, which are small at high collision energies, and the quantum corrections to Boltzmann statistics, which are suppressed due to $m_c/T\gg 1$. Consequently, charmonium states are enhanced by a factor of $\lambda_c^2$, analogous to the enhancement by $g_c^2$ in the hadronic formulation of the SHM.  Note that no excluded volume corrections are applied in Equation (\ref{eq:lambda_c}), as quarks are thought to be true point particles.

\section{Comparison of $\boldsymbol{J/\psi}$ multiplicities}
\label{sec:Comparison of multiplicities}

Lattice calculations have demonstrated that $J/\psi $ states already exist in the QGP  up to at least 1.5 $T_c$ \cite{latticeQCD03,Karsch03,latticeQCD07}. This is plausible as the $J/\psi $ are a strongly bound and compact
colour singlet state. We now assume that these states do exist prior to hadronization and are produced sufficiently fast (perturbatively via e.g. $J/\psi + g \leftrightarrow c + \bar{c} $ or by a nonperturbative mechanism) to obey thermal equilibrium under the constraint of conserving the total number of charm quarks.
This scenario of $J/\psi $ equilibration and saturation in the QGP phase just prior the crossover regime  is then the analogue of the SHM picture of charmonium production after the hadronization. 
Hence, following this reasoning and the considerations from the previous sections, two different equations for the multiplicity of the $J/\psi$ meson $N_{J/\psi}$ apply:
\begin{align}
N_{J/\psi}^{\mathrm{SHM}}=\frac{3 V_{\mathrm{SHM}}}{2\pi^2}\,g_c^2\int_{0}^{\infty}\!\mathrm{d}p\frac{p^2}{\exp\left(\frac{E_{J/\psi}-\mu_{J/\psi}}{T}\right)-1}\label{eq:J/psi SHM}
\end{align}
and
\begin{align}
N_{J/\psi}^{\mathrm{QGP}}=\frac{3 V_{\mathrm{QGP}}}{2\pi^2}\,\lambda_c^2\int_{0}^{\infty}\!\mathrm{d}p\frac{p^2}{\exp\left(\frac{E_{J/\psi}-\mu_{J/\psi}}{T}\right)-1}\label{eq:J/psi QGP} \, .
\end{align}
Here the equivalent integral form \cite{And05} of Equation (\ref{eq:multiplicity}) specifically for the $J/\psi$ meson is used. Both approaches should yield compatible results, as the deviation of charmed hadrons and charmonia from chemical equilibrium in the hadron gas phase should be governed by the corresponding deviation of the constituent charm quarks from the chemical equilibrium in the QGP. Because the acceptable range for $T$ is fixed by experimental data as well as theoretical calculations, the conditions for the remaining parameters, especially for the charm quark mass $m_c$, can be investigated.\\

\begin{figure}[htb]
\centering
\includegraphics[width=\linewidth]{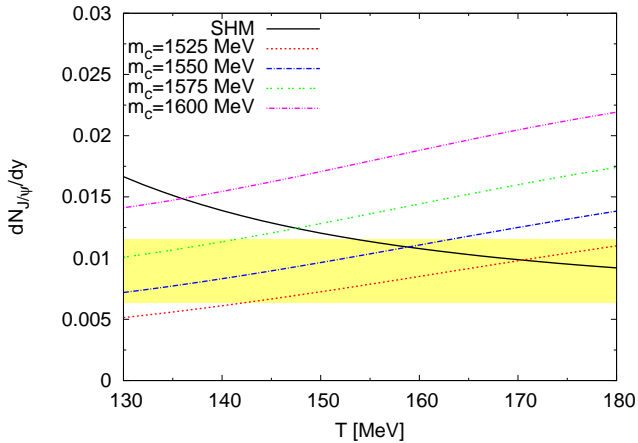}
\caption{Comparison of $J/\psi$ yields in the hadronic and the partonic formulation. High sensitivity to the value chosen for $m_c$ is evident in the results. The input parameters are fixed to $N_{\mathrm{ch}}=730$ and $N_{c\bar{c}}^{\mathrm{dir}}=1.92$, which were used to reproduce particle yields from RHIC data in \citep{And03}. The shaded band represents the midrapidity yield reported by PHENIX \cite{PHENIX06}.}
\label{fig:mass_scaling}
\end{figure} 
The resulting $J/\psi$ yields according to Equations (\ref{eq:J/psi SHM}) and (\ref{eq:J/psi QGP}) are shown in Figure \ref{fig:mass_scaling} for a fixed number of $c\bar{c}$-pairs $N_{c\bar{c}}^{\mathrm{dir}}=1.92$, which is consistent with the lower bounds for the charm production cross-section reported by the PHENIX collaboration \cite{Adare:2006hc,Adare:2008ac} for four different values of $m_c$, including feed-down from higher hadronic states. The two approaches show different dependences on the freeze-out temperature, decreasing slightly over the whole interval for higher values in the hadronic case while consistently increasing in the partonic case. While the $J/\psi$ multiplicity is only mildly dependent on the temperature in both cases, dependence on the effective dressed charm quark mass used in Equation (\ref{eq:lambda_c}) is strong in the partonic formulation, getting more pronounced for higher masses as evident by the increasing displacement of the curves for a constant mass difference.

The $J/\psi$ color singlet states which were formed in the plasma equilibrate with the medium under the constraint of detailed balance
are compact and strongly bound charmonium states and should, arguably, survive any phenomenon occurring at hadronization, therefore constituting the resulting, measurable yield in the hadronic phase. The principle of  detailed balance in forming
and annihilating such states, thus providing their thermal yields under the constraint of floating charm quarks or $D$ meson-like states, should also hold during the nonperturbative hadronization process in the crossover regime.
Then both approaches need  to give approximately the same $J/\psi$ multiplicities for a given freeze-out temperature. 
From the above comparison this condition favours an effective dressed mass of $m_c\simeq 1550\,\mathrm{MeV}$, updating the value of $m_c\simeq 1.6$-$1.7\,\mathrm{GeV}$ from \cite{Grandchamp:2003uw} and verifying the proposed approach with recent charm measurements from PHENIX.\cite{PHENIX06}\\

\begin{figure}[htb]
\centering
\includegraphics[width=\linewidth]{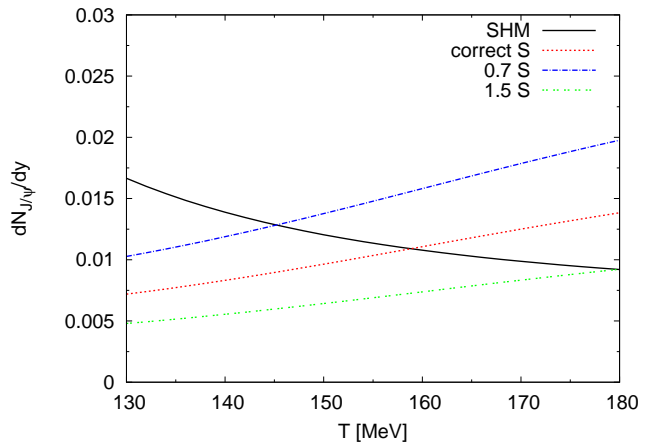}
\caption{To illustrate the influence of the exact value for the entropy used in the calculation, $J/\psi$ yields for scaled values for the entropy of the hadronic fireball are shown here compared to the results of the standard SHM formulation. A charm quark mass of $m_c=1550\,\mathrm{MeV}$ was used.}
\label{fig:entropy_scaling}
\end{figure}

The sensitivity of the $J/\psi$ multiplicities on potential uncertainties in calculating the total entropy density in
the hadronic gas is shown in Figure \ref{fig:entropy_scaling}. Counter-intuitively, a higher volume decreases the final yield, due to the inverse proportionality of $\lambda_c$ to $V_{\mathrm{QGP}}$ as shown in Equation (\ref{eq:lambda_c}), which in turn contributes quadratically to the multiplicity. There is only weak sensitivity to this parameter and therefore to the assumption of exact conservation made in Equation~(\ref{eq:V_QGP}).

\section{Conclusions and Outlook}
\label{sec:Conclusions}

Using a complete Statistical Hadronization Model on the hadronic phase and adapting the formalism for the charm fugacity to work on the partonic phase, it was possible to compare $J/\psi$ production in both pictures under the assumption of entropy conservation across the hadronization process. The yields in the partonic picture show a mild dependence on the freeze-out temperature similar to SHM results over the investigated temperature range. The in-medium charm quark mass required for compatible results in both descriptions can be determined for any value of the freeze-out temperature, which is limited to a certain interval due to existing experimental data. Therefore, limits for the effective dressed mass of the charm quark can be obtained from this approach. The results suggest a very narrow interval for $m_c$ centered at about $1550\,\mathrm{MeV}$ in order to obtain compatible results for both formulations of the model and experimental data at sensible values for $T$. \\

While this value is of course subject to the inaccuracies introduced by the various approximations and assumptions made, it is decisively closer to the vacuum mass of the charm quark at $1275\,\mathrm{MeV}$ \cite{PDG} and also results from lattice QCD which suggest $m_c=1348\,\mathrm{MeV}$ \cite{Carrasco:2014cwa} than the favoured mass from \cite{Grandchamp:2003uw} of $m_c=1.6$-$1.7\,\mathrm{GeV}$ or the popular approach of substituting $m_c$ with the $D$ meson mass $m_D=1870\,\mathrm{MeV}$ \cite{Liu:2009nb,Zhuang:2007zz,Zhou:2014kka}. In \cite{Zhou:2014kka} in particular, the authors have solved  temporal evolution equations for $J/\psi $ states in the QGP, and their results are in agreement with the various data at RHIC and LHC energies. As an outlook we will further test our result by analyzing the behaviour of charm quarks and the $J/\psi$ in the QGP via kinetic processes implemented in \cite{Uphoff:2011fu}  in the partonic transport model BAMPS (\emph{Boltzmann Approach of MultiParton Scatterings}) \cite{BAMPS05,BAMPS07}, which simulates the 3+1 space-time evolution of the QGP and then  to compare to the results in \cite{Zhou:2014kka}.
\section*{Acknowledgements}
We would like to thank A.\ Andronic, R. Rapp, Y. Liu, and H. van Hees for helpful discussions. This work was supported by the Bundesministerium f\"ur Bildung und Forschung (BMBF) and by the Helmholtz International Center for FAIR within the framework of the LOEWE program launched by the State of Hesse.
\bibliography{ref}
\end{document}